\documentclass[aps,preprint,showpacs]{revtex4}%
\usepackage{amsfonts}
\usepackage{amsmath}
\usepackage{amssymb}
\usepackage{graphicx}%

\begin{document}
\title{Dynamics of Dipolar Spinor Condensates}
\author{Rong Cheng, J.-Q. Liang, Yunbo Zhang$^*$}
\affiliation{Department of Physics and Institute of Theoretical Physics, Shanxi University,
Taiyuan 030006, P. R. China}
\date{\today}

\begin{abstract}
We study the semiclassical dynamics of a spinor condensate with the magnetic
dipole-dipole interaction included. The time evolution of the population
imbalance and the relative phase among different spin components depends
greatly on the relative strength of interactions as well as on the initial
conditions. The interplay of spin exchange and dipole-dipole interaction makes
it possible to manipulate the atomic population on different components,
leading to the phenomena of spontaneous magnetization and Macroscopic Quantum
Self Trapping. Simple estimate demonstrates that these effects are accessible
and controllable by modifying the geometry of the trapping potential.

\end{abstract}

\maketitle

\section{Introduction}

One of the current experimental advances of considerable importance in the
context of ultracold atoms is the demonstration of multicomponent
Bose-Einstein condensate (BEC). In particular several groups have created
spinor condensates of $^{23}$Na atoms \cite{1} and $^{87}$Rb atoms \cite{2} by
transferring spin polarized condensates into far-off-resonance optical dipole
traps, where the spin degree of freedom becomes active and the spinor nature
of the condensates is manifested. The spin-mixing dynamics and its dependence
on the magnetic field have been investigated in detail for both spin-1 and
spin-2 condensates in a number of theoretical \cite{3} and experimental works
\cite{1,2}. Since the spin degree of freedom becomes accessible in an optical
trap, the magnetic dipole-dipole interactions (MDDI) which arise from
intrinsic or induced field of magnetic dipole moment \cite{4,5} should be
taken into account in addition to the isotropic $s$-wave contact interaction.
Due to their long-range and vectorial characters, the MDDI may largely enrich
the variety of phenomena observed in condensates. Therefore this opens up new
directions of research, such as the ground-state structure of spin-1 dipolar
condensates in a single trap \cite{6} as well as the ground-state magnetic
properties of spinor BEC confined in deep optical lattices \cite{7,8,9}, where
only the dipolar interactions between condensates at different sites are
considered. Recently, a beautiful experiment in Stuttgart \cite{10} has
demonstrated BEC in a gas of chromium atoms $^{52}Cr$, the MDDI of which is a
factor of 36 higher than that for alkali atoms. This achievement makes
possible the studies of the anisotropic long-range interactions in degenerate
quantum gases.

The present paper exploits semiclassical dynamics of spin-1 dipolar
condensates in a single trap and shows the time evolutions of the population
imbalance and the relative phase among different spin components \cite{6,11}.
This study may help us gain some insights into the properties of this dipolar
spinor condensate, such as quantum phase diffusion, spontaneous magnetization
\cite{12} and Macroscopic Quantum Self Trapping (MQST). Especially, MQST known
as a novel nonlinear effect has already been predicted theoretically \cite{13}
and very recently observed experimentally in double-well trap \cite{14} and
periodic optical lattice \cite{15}. The self trapping effect sustains a
self-maintained\ interwell population imbalance during the
nonlinear\textbf{\ }tunneling process. Here we will show that for a dipolar
condensate the interplay between spin-exchange and MDDI gives rise to the
intercomponent MQST naturally. Moreover, due to the adjustability of the MDDI,
the experimental observation of the MQST effect in dipolar spinor condensates
can be\textbf{\ }expected\textbf{.}

The paper is organized as follows. In Sec. II we describe the dipolar spinor
BEC model and derive the equations of motion of semiclassical dynamics in
terms of symmetry. Sec. III is devoted to the properties of the equations of
motion such as spontaneous magnetization and spin-mixing dynamics. The time
evolutions of the population imbalance and the relative phase among different
spin components are investigated and quantum tunneling and self trapping among
different spin components are examined carefully. Finally, a brief summary is
given in Sec. IV.

\section{ The Model and the Equations of Motion}

We consider a spin $F=1$ dipolar condensate with $N$ bosons. In second
quantized form, the Hamiltonian $\widehat{H}_{tot}$ subject to both
spin-exchanging collisions $\widehat{H}_{sp}$ and MDDI $\widehat{H}_{dd}$
reads \cite{6}
\[
\widehat{H}_{tot}=\widehat{H}_{sp}+\widehat{H}_{dd}%
\]
with
\begin{align}
\widehat{H}_{sp}  &  =\int d\mathbf{r}\widehat{\psi}_{\alpha}^{\dagger}\left(
\mathbf{r}\right)  \left[  -\frac{\bigtriangledown^{2}}{2M}+V_{T}\left(
\mathbf{r}\right)  \right]  \widehat{\psi}_{\alpha}\left(  \mathbf{r}\right)
\nonumber\\
&  +\frac{c_{0}}{2}\int d\mathbf{r}\widehat{\psi}_{\alpha}^{\dagger}\left(
\mathbf{r}\right)  \widehat{\psi}_{\beta}^{\dagger}\left(  \mathbf{r}\right)
\widehat{\psi}_{\alpha}\left(  \mathbf{r}\right)  \widehat{\psi}_{\beta
}\left(  \mathbf{r}\right) \nonumber\\
&  +\frac{c_{2}}{2}\int d\mathbf{r}\widehat{\psi}_{\alpha}^{\dagger}\left(
\mathbf{r}\right)  \widehat{\psi}_{\beta}^{\dagger}\left(  \mathbf{r}\right)
\mathbf{F}_{\alpha\alpha^{\prime}}\cdot\mathbf{F}_{\beta\beta^{\prime}%
}\widehat{\psi}_{\beta^{\prime}}\left(  \mathbf{r}\right)  \widehat{\psi
}_{\alpha^{\prime}}\left(  \mathbf{r}\right)  \label{1}%
\end{align}
and
\begin{align}
\widehat{H}_{dd}  &  =\frac{c_{d}}{2}\int d\mathbf{r}\int d\mathbf{r}^{\prime
}\frac{1}{\left\vert \mathbf{r}-\mathbf{r}^{\prime}\right\vert ^{3}%
}\nonumber\\
&  \times{\Huge [}\widehat{\psi}_{\alpha}^{\dagger}\left(
\mathbf{r}\right) \widehat{\psi}_{\beta}^{\dagger}\left(
\mathbf{r}^{\prime}\right)
\mathbf{F}_{\alpha\alpha^{\prime}}\cdot\mathbf{F}_{\beta\beta^{\prime}%
}\widehat{\psi}_{\beta^{\prime}}\left(  \mathbf{r}\right)  \widehat{\psi
}_{\alpha^{\prime}}\left(  \mathbf{r}^{\prime}\right) \nonumber\\
&  -3\widehat{\psi}_{\alpha}^{\dagger}\left(  \mathbf{r}\right)  \widehat
{\psi}_{\beta}^{\dagger}\left(  \mathbf{r}^{\prime}\right)  \left(
\mathbf{F}_{\alpha\alpha^{\prime}}\cdot\mathbf{e}\right)  \left(
\mathbf{F}_{\beta\beta^{\prime}}\cdot\mathbf{e}\right)  \widehat{\psi}%
_{\beta^{\prime}}\left(  \mathbf{r}\right)  \widehat{\psi}_{\alpha^{\prime}%
}\left(  \mathbf{r}^{\prime}\right)  {\Huge ]} , \label{2}%
\end{align}
where $\widehat{\psi}_{\alpha}\left(  \mathbf{r}\right)  $ $\left(
\alpha=0,\pm1\right)  $ are the field annihilation operators for an atom in
the hyperfine state $|F=1,m_{F}=\alpha\rangle.$ The hyperfine spin
$\mathbf{F}$ of one single atom is expressed in the spin-1 matrix
representation, and $M$ is the mass of the atom. A summation over repeated
indices is assumed in Eqs. (\ref{1}) and (\ref{2}). The external trapping
potential $V_{T}\left(  \mathbf{r}\right)  $ is spin independent for a far
off-resonant optical dipole trap which makes atomic spin degree of freedom
completely accessible. The two coefficients $c_{0}=4\pi\hbar^{2}\left(
a_{0}+2a_{2}\right)  /3M$, $c_{2}=4\pi\hbar^{2}\left(  a_{2}-a_{0}\right)
/3M$ characterize the density-density and spin-spin collisional interactions,
respectively, with $a_{f}$ $\left(  f=0,2\right)  $ being the $s$-wave
scattering length for two spin-1 atoms in the combined symmetric channel of
total spin $f.$ And the dipolar interaction parameter is $c_{d}=\mu_{0}\mu
_{B}^{2}g_{F}^{2}/4\pi$ with $g_{F}$ being the Land\'{e} $g$-factor, $\mu_{B}$
the Bohr magneton. $\mathbf{e}=\left(  \mathbf{r}-\mathbf{r}^{\prime}\right)
/\left\vert \mathbf{r}-\mathbf{r}^{\prime}\right\vert $ is a unit vector. For
the two experimentally created spinor condensates ($^{23}$Na and $^{87}$Rb),
we have $\left\vert c_{2}\right\vert \ll c_{0}.$

For a condensate with its symmetry axis chosen to be along the quantization
axis $z$ (which happens to be the most experimentally relevant cases),
$\widehat{H}_{dd}$ takes a very simple form as shown in Ref. \cite{6}. Under
the single mode approximation (SMA), namely, $\widehat{\psi}_{\alpha}\left(
\mathbf{r}\right)  \approx\phi\left(  \mathbf{r}\right)  \widehat{a}_{\alpha
},$ where $\phi\left(  \mathbf{r}\right)  $ is the spin-independent condensate
spatial wave function, $\widehat{a}_{\alpha}$ is the annihilation operator for
$m_{F}=\alpha$ component, the total Hamiltonian apart from a trivial term is
given as
\begin{align}
\widehat{H}_{tot}  &  =\widehat{H}_{sp}+\widehat{H}_{dd}=\varepsilon
\sum_{\alpha}\widehat{a}_{\alpha}^{\dagger}\widehat{a}_{\alpha}+U_{0}%
\sum_{\alpha,\beta}\widehat{a}_{\alpha}^{\dagger}\widehat{a}_{\beta}^{\dagger
}\widehat{a}_{\beta}\widehat{a}_{\alpha}\nonumber\\
&  +U_{2}(\widehat{a}_{1}^{\dagger}\widehat{a}_{1}^{\dagger}\widehat{a}%
_{1}\widehat{a}_{1}+\widehat{a}_{-1}^{\dagger}\widehat{a}_{-1}^{\dagger
}\widehat{a}_{-1}\widehat{a}_{-1}-2\widehat{a}_{1}^{\dagger}\widehat{a}%
_{-1}^{\dagger}\widehat{a}_{1}\widehat{a}_{-1}\nonumber\\
&  +2\widehat{a}_{1}^{\dagger}\widehat{a}_{0}^{\dagger}\widehat{a}_{1}%
\widehat{a}_{0}+2\widehat{a}_{-1}^{\dagger}\widehat{a}_{0}^{\dagger}%
\widehat{a}_{-1}\widehat{a}_{0}+2\widehat{a}_{0}^{\dagger}\widehat{a}%
_{0}^{\dagger}\widehat{a}_{1}\widehat{a}_{-1}+2\widehat{a}_{1}^{\dagger
}\widehat{a}_{-1}^{\dagger}\widehat{a}_{0}\widehat{a}_{0})\nonumber\\
&  +U_{d}(2\widehat{a}_{1}^{\dagger}\widehat{a}_{1}^{\dagger}\widehat{a}%
_{1}\widehat{a}_{1}+2\widehat{a}_{-1}^{\dagger}\widehat{a}_{-1}^{\dagger
}\widehat{a}_{-1}\widehat{a}_{-1}-4\widehat{a}_{1}^{\dagger}\widehat{a}%
_{-1}^{\dagger}\widehat{a}_{1}\widehat{a}_{-1}\nonumber\\
&  -2\widehat{a}_{1}^{\dagger}\widehat{a}_{0}^{\dagger}\widehat{a}_{1}%
\widehat{a}_{0}-2\widehat{a}_{-1}^{\dagger}\widehat{a}_{0}^{\dagger}%
\widehat{a}_{-1}\widehat{a}_{0}-2\widehat{a}_{0}^{\dagger}\widehat{a}%
_{0}^{\dagger}\widehat{a}_{1}\widehat{a}_{-1}\nonumber\\
&  -2\widehat{a}_{1}^{\dagger}\widehat{a}_{-1}^{\dagger}\widehat{a}%
_{0}\widehat{a}_{0}+\widehat{a}_{1}^{\dagger}\widehat{a}_{1}+\widehat{a}%
_{-1}^{\dagger}\widehat{a}_{-1}-2\widehat{a}_{0}^{\dagger}\widehat{a}_{0}),
\label{3}%
\end{align}
where $\varepsilon=\int d\mathbf{r}\phi^{\ast}\left(  \mathbf{r}\right)
[-\bigtriangledown^{2}/2M+V_{T}\left(  \mathbf{r}\right)  ]\phi\left(
\mathbf{r}\right)  $ is assumed to be of the same value for all components.
$U_{0,2}=\left(  c_{0,2}/2\right)  \int\left\vert \phi\left(  \mathbf{r}%
\right)  \right\vert ^{4}d\mathbf{r}$ characterizes the spin-changing
collisions and $U_{d}=$ $\left(  c_{d}/4\right)  \int\int d\mathbf{r}%
d\mathbf{r}^{\prime}\left\vert \phi\left(  \mathbf{r}\right)  \right\vert
^{2}\left\vert \phi\left(  \mathbf{r}^{\prime}\right)  \right\vert ^{2}\left(
1-3\cos^{2}\theta_{e}\right)  /\left\vert \mathbf{r}-\mathbf{r}^{\prime
}\right\vert ^{3}$ denotes the MDDI with $\theta_{e}$ being the polar angle of
$\left(  \mathbf{r}-\mathbf{r}^{\prime}\right)  $. Before we proceed to
examine the semiclassical dynamics of $\widehat{H}_{tot},$ we would like to
emphasize that $\widehat{H}_{sp}$ is invariant under spin rotation, i.e., it
possesses a $SO(3)$\ symmetry in spin space \cite{16}. The presence of the
MDDI breaks this symmetry into $SO(2)$,\ which means $\widehat{H}_{tot}$\ has
only axial symmetry in spin space as we chosen the quantization axis along
$z$. In this sense the MDDI plays a similar role to that of an external
magnetic field.

In the mean-field theory the condensate is usually considered to be in a
coherent state \cite{11}%

\begin{equation}
|\overrightarrow{z}\rangle=\exp\left(  -\frac{1}{2}\sum_{\alpha}\left\vert
z_{\alpha}\right\vert ^{2}\right)  \exp\left(  \sum_{\alpha}z_{\alpha}%
\widehat{a}_{\alpha}^{\dagger}\right)  |0\rangle, \label{4}%
\end{equation}
where $|0\rangle$ is the vacuum state. The complex numbers $z_{\alpha}$ are
nothing but the macroscopic wave functions for the atoms in the hyperfine
level $|F=1,m_{F}=\alpha\rangle$ with population $N_{\alpha}$ and phase
$\theta_{\alpha},$ i.e.,
\begin{equation}
z_{\alpha}=\sqrt{N_{\alpha}}e^{-i\theta_{\alpha}}. \label{5}%
\end{equation}
The time-dependent variational principle of the system
\begin{equation}
\delta S=\delta\int i\hbar\langle\overrightarrow{z}|\overset{.}%
{\overrightarrow{z}}\rangle-\langle\overrightarrow{z}|H|\overrightarrow
{z}\rangle dt=0 \label{6}%
\end{equation}
gives\ rise to the Hamiltonian equations of motion in complex coordinates
\begin{equation}
i\hbar\overset{\cdot}{z}_{\alpha}=\frac{\partial H_{0}}{\partial z_{\alpha
}^{\ast}},\text{ }-i\hbar\overset{\cdot}{z}_{\alpha}^{\ast}=\frac{\partial
H_{0}}{\partial z_{\alpha}} \label{7}%
\end{equation}
with
\begin{align}
H_{0}\left(  N_{\alpha},\theta_{\alpha}\right)   &  =\left(  \varepsilon
+U_{d}\right)  N-3U_{d}N_{0}+U_{0}N^{2}+\left(  U_{2}+2U_{d}\right)  \left(
N_{1}-N_{-1}\right)  ^{2}\nonumber\\
&  +2\left(  U_{2}-U_{d}\right)  N_{0}[N_{1}+N_{-1}+2\sqrt{N_{1}N_{-1}}%
\cos\left(  2\theta_{0}-\theta_{1}-\theta_{-1}\right)  ]. \label{8}%
\end{align}
Note that the dynamics governed by Hamiltonian (\ref{3}) conserves the total
atom numbers $N=\sum_{\alpha}N_{\alpha}$ and the total hyperfine spin of the
condensate in the direction of the quantization axis\textsl{ }$I=\left(
N_{1}-N_{-1}\right)  \hbar.$ In terms of the following canonical variables
\begin{equation}%
\begin{array}
[c]{l}%
\varphi_{1}=\frac{\left(  \theta_{1}+\theta_{0}+\theta_{-1}\right)  }{3}\\
\varphi_{2}=\theta_{0}-\frac{\left(  \theta_{1}+\theta_{-1}\right)  }{2}\\
\varphi_{3}=\theta_{1}-\theta_{-1}%
\end{array}
\text{ and }%
\begin{array}
[c]{l}%
\Omega_{1}=N_{1}+N_{0}+N_{-1}\\
\Omega_{2}=\frac{2}{3}N_{0}-\frac{1}{3}\left(  N_{1}+N_{-1}\right) \\
\Omega_{3}=\frac{\left(  N_{1}-N_{-1}\right)  }{2}%
\end{array}
\label{9}%
\end{equation}
the Hamiltonian (\ref{8}) becomes cyclic in the coordinates $\varphi_{1}$ and
$\varphi_{3}$
\begin{align}
&  H_{0}\left(  \Omega_{1},\Omega_{2},\Omega_{3},\varphi_{2}\right)
=\varepsilon\Omega_{1}+U_{0}\Omega_{1}^{2}-3U_{d}\Omega_{2}\nonumber\\
&  +4\left(  U_{2}+2U_{d}\right)  \Omega_{3}^{2}+2\left(  U_{2}-U_{d}\right)
\nonumber\\
&  \times{\Huge [}\left(  \frac{1}{3}\Omega_{1}+\Omega_{2}\right)  \left(
\frac{2}{3}\Omega_{1}-\Omega_{2}\right) \nonumber\\
&  +\sqrt{\left(  \frac{2}{3}\Omega_{1}-\Omega_{2}\right)  ^{2}-4\Omega
_{3}^{2}}\left(  \frac{1}{3}\Omega_{1}+\Omega_{2}\right)  \cos2\varphi
_{2}{\Huge ]}. \label{10}%
\end{align}
Two important consequences follow as a result of the cyclic coordinates
$\varphi_{1}$ and $\varphi_{3}.$ Firstly the mean value of the total number of
atoms $N=\Omega_{1}$ and that of the total hyperfine spin of the condensate
$I=2\hbar\Omega_{3}$ are constants of motion. The dynamics, on the other hand,
involves only one pair of variables $\left\{  \varphi_{2},\Omega_{2}\right\}
$ as
\[
\dot{\varphi}_{2}=\frac{\partial H_{0}}{\partial\Omega_{2}},\text{\qquad}%
\dot{\Omega}_{2}=-\frac{\partial H_{0}}{\partial\varphi_{2}},
\]
Define $\xi_{i}=\Omega_{i}/N$ $\left(  i=1,2,3\right)  $ and $\eta
=\sqrt{(\frac{2}{3}-\xi_{2})^{2}-4\xi_{3}^{2}}$, we obtain%
\begin{align}
\hbar\dot{\xi}_{2}  &  =4N\left(  U_{2}-U_{d}\right)  \eta\left(  \frac{1}%
{3}+\xi_{2}\right)  \sin2\varphi_{2}\nonumber\\
-\hbar\dot{\varphi}_{2}  &  =3U_{d}+2N\left(  U_{2}-U_{d}\right)  \left(
2\xi_{2}-\frac{1}{3}\right) \nonumber\\
&  +2N\left(  U_{2}-U_{d}\right)  \frac{\left(  \frac{2}{3}-\xi_{2}\right)
\left(  2\xi_{2}-\frac{1}{3}\right)  +4\xi_{3}^{2}}{\eta}\cos2\varphi_{2}.
\label{11}%
\end{align}
That means not only $\xi_{2}$ but also $\xi_{3}$ are involved in the dynamics.

The variables $\xi_{2}$ and $\varphi_{2}$ are canonically conjugate in a
classical Hamiltonian
\begin{align}
H  &  =-\frac{\left(  U_{2}-U_{d}\right)  N}{2}\left(  2\xi_{2}-\frac{1}%
{3}\right)  ^{2}\nonumber\\
&  +2\left(  U_{2}-U_{d}\right)  N\eta\left(  \frac{1}{3}+\xi_{2}\right)
\cos2\varphi_{2}-3U_{d}\xi_{2}. \label{12}%
\end{align}
In a simple mechanical analogy, $H$ describes a nonrigid pendulum. The Bose
Josephson junction\ (BJJ) tunneling current between different spin components
is given by
\begin{equation}
I_{BJJ}=\frac{\dot{\xi}_{2}N}{2}=I_{0}\eta\left(  \frac{1}{3}+\xi_{2}\right)
\sin2\varphi_{2}, \label{13}%
\end{equation}
where $I_{0}=2N^{2}\left(  U_{2}-U_{d}\right)  $. It differs from BJJ
tunneling current of two weakly linked BEC in a double-well potential in its
further nonlinearity in $\xi_{2}$ \cite{13}.

\section{Properties of the Equations of Motion}

\subsection{Spontaneous magnetization}

We firstly study the equilibrium configurations of the system. They are
determined by the classical equations of motion (\ref{11}) after setting the
time derivative terms to zero%
\begin{align}
0  &  =\eta\left(  \frac{1}{3}+\xi_{2}\right)  \sin2\varphi_{2}\label{14}\\
0  &  =\Lambda\pm{\Huge [}2\xi_{2}-\frac{1}{3}\nonumber\\
&  +\frac{\left(  \frac{2}{3}-\xi_{2}\right)  \left(  2\xi_{2}-\frac{1}%
{3}\right)  +4\xi_{3}^{2}}{\eta}\cos2\varphi_{2}{\Huge ]}. \label{15}%
\end{align}
We have defined a dimensionless parameter $\Lambda=\frac{3U_{d}}{2\left\vert
U_{2}-U_{d}\right\vert N}$ characterizing the relative interaction strength of
the MDDI. In Eq. (\ref{15}) plus sign corresponds to the case of $\left(
U_{2}-U_{d}\right)  >0,$\ while minus sign corresponds to $\left(  U_{2}%
-U_{d}\right)  <0$\ (In the following we take the former case as an
example)$.$ From the definition of $\eta$ we know that the constant of motion
$\xi_{3}$ takes values in the interval $-\frac{1}{2}<\xi_{3}<\frac{1}{2}$ and
the dynamic variable $\xi_{2}$ in the interval $-\frac{1}{3}<\xi_{2}<\frac
{2}{3}-2\left\vert \xi_{3}\right\vert .$ In the following discussion about the
solutions of the equilibrium equations we separately consider three cases in
order to see more clearly how these solutions depend on the relative phase
$\varphi_{2}$.

(1) Equilibrium configuration with $\cos2\varphi_{2}=1$ (or $2\varphi
_{2}=2k\pi$)$.$

The equilibrium value of $\xi_{2}$ is thus given by equation (\ref{15}) with
$\cos2\varphi_{2}=1,$ which has only one solution in the interval
$-\infty<\Lambda<1+\sqrt{1-\left(  2\xi_{3}\right)  ^{2}}.$ When
$\Lambda\rightarrow-\infty$ the equilibrium value of $\xi_{2}$ approaches its
upper boundary $\frac{2}{3}-2\left\vert \xi_{3}\right\vert $. In this case the
fractions of atoms $n_{\alpha}=N_{\alpha}/N$ ($\alpha=0,\pm1$) occupying three
hyperfine states are $n_{1}=\left\vert \xi_{3}\right\vert +\xi_{3},$
$n_{0}=1-2\left\vert \xi_{3}\right\vert ,$ $n_{-1}=\left\vert \xi
_{3}\right\vert -\xi_{3}$ $.$ On the other hand when $\Lambda=1+\sqrt
{1-\left(  2\xi_{3}\right)  ^{2}},$ $\xi_{2}$ is at the lower boundary
$-\frac{1}{3},$ where the fractions of atoms occupying the hyperfine states
are $n_{1}=\frac{1}{2}\left(  1+2\xi_{3}\right)  ,$ $n_{0}=0,$ $n_{-1}%
=\frac{1}{2}\left(  1-2\xi_{3}\right)  .$ Besides, when $\xi_{3}=0$ the
equilibrium value of $\xi_{2}$ is $\xi_{2}=\frac{1}{2}\left(  \frac{1}%
{3}-\Lambda\right)  $ in the interval $-2<\Lambda<2$ and the occupation
fractions of three hyperfine states are respectively $n_{0}=\frac{1}{2}\left(
1-\Lambda\right)  ,$ $n_{1}=n_{-1}=\frac{1}{4}\left(  1+\Lambda\right)  .$

(2) Equilibrium configuration with $\cos2\varphi_{2}=-1$ (or $2\varphi
_{2}=\left(  2k+1\right)  \pi$)$.$

Now the equilibrium value of $\xi_{2}$ is given by equation (\ref{15}) with
$\cos2\varphi_{2}=-1.$ Similarly it has again only one solution in the
interval $\Lambda>1-\sqrt{1-\left(  2\xi_{3}\right)  ^{2}}.$ When $\Lambda=$
$1-\sqrt{1-\left(  2\xi_{3}\right)  ^{2}},$ $\xi_{2}$ is at the lower boundary
$-\frac{1}{3}.$ In particular, when $\xi_{3}=0$ the solution only exists for
vanished dipole-dipole interaction $U_{d}$.

(3) Equilibrium configuration with $\cos2\varphi_{2}=0$ (or Equilibrium
configuration which does not depend on the phase $\varphi_{2}$)\textsl{.}

In this case there exist two solutions which are independent of the value of
$\Lambda.$ One solution appears at the lower boundary $\xi_{2}=-\frac{1}{3}$
and the occupation fractions are $n_{1}=n_{-1}=\frac{1}{2},n_{0}=0.$ The other
one appears at the upper boundary $\xi_{2}=\frac{2}{3}-2\left\vert \xi
_{3}\right\vert $ and the occupation fractions are $n_{1}=\left\vert \xi
_{3}\right\vert +\xi_{3},$ $n_{0}=1-2\left\vert \xi_{3}\right\vert ,$
$n_{-1}=\left\vert \xi_{3}\right\vert +\xi_{3}.$ Interestingly when $\xi
_{3}=0$ the occupation fractions become $n_{1}=n_{-1}=0$ and $n_{0}=1,$
corresponding to the \textquotedblleft polar\textquotedblright\ state of the condensate.

The mean field theory predicted a polar ground state of spinor BEC for $^{23}%
$Na in which the atoms interact antiferromagnetically with each other in the
absence of the MDDI \cite{16}. The intrinsic magnetic moment of particles may,
however, contribute together to form a Bose-Einstein Ferromagnetism (BEF)
\cite{12}. Spinor bosons carry magnetic moments $\mu$ and $M=\mu\left\vert
\psi_{1}\right\vert ^{2}$ represents the magnetization. Spontaneous
magnetization means\ that $M$\ remains finite even if the external magnetic
field $B$ vanishes, namely, the ground state of spinor BEC is always
ferromagnetic state. Our calculations prefer the latter scenario.
Specifically, above results suggest that for most given values of $\Lambda
$\ and initial relative phase among the three components, when $B=0$\ and
$U_{2}>0,$\ we have $M$\ $\neq0$\ , i.e., the system is magnetized
spontaneously once the dipolar spinor BEC is realized. As mentioned in Ref.
\cite{12}, one can provide a direct way of confirming spontaneous
magnetization experimentally. Considering that both the sign and the magnitude
of the MDDI coefficient $c_{d}$ depend on the geometric shape of the
condensate \cite{4,10,17,18}, we can conveniently adjust the trap aspect ratio
to manipulate the magnetic property of its ground state.

\subsection{Spin-Mixing Dynamics}

A key feature of dipolar spinor BEC is that besides the usual two-body
repulsive hard-core interactions, there also exist spin-exchange interaction
and MDDI which lead to spin mixing within the condensates. Population can be
transferred from one spin state to another under internal nonlinear
interactions without the presence of external fields. Insight into the complex
dynamics of our system can be gained by employing the method of action-angle
variables \cite{19}. In the subsequent section we present some numerical
results and show the time evolutions of the population imbalance and the
relative phase among different spin components for two cases a) $\xi_{3}=0$
and b) $\xi_{3}\neq0,$ i.e., whether total hyperfine spin of the condensate is
zero or not. Time has been rescaled in units of $2\left\vert U_{2}%
-U_{d}\right\vert N / \hbar $ in Figures 1 and 2.

Figure 1 shows solutions of the population imbalance and relative phase of
Eqs. (\ref{11}) with initial conditions $\xi_{2}\left(  0\right)  =0.12,$
$\xi_{3}\left(  0\right)  =0.25,$ $\varphi_{2}\left(  0\right)  =\frac{\pi}%
{2}$ for the relative interaction parameters $\Lambda=\frac{3U_{d}%
}{2\left\vert U_{2}-U_{d}\right\vert N}$ $=0.00,0.40,0.60$ and $0.75$,
respectively. The left column exhibits the time evolutions of the population
imbalance between $0$ and $\pm1$ components. Josephson sinusoidal oscillations
are observed as $\Lambda$\ increases in Figs. 1a, 1b. In Fig. 1c, there is a
critical transition point for $\Lambda=\Lambda_{c}=0.60$. Increasing $\Lambda
$\ further for example to $\Lambda=0.75$,\ the population of\ each equivalent
component oscillates\ around a nonzero time averaged value, which gives a net
population imbalance $\left\langle \xi_{2}\left(  t\right)  \right\rangle >0$.
This is closely related to the MQST phenomenon. Simultaneously, the right
column shows the time evolutions of the relative phase between $0$ and $\pm1$
components and indicates that as $\Lambda$ increases $\varphi_{2}\left(
t\right)  $ varies from a monotonically increasing function of time to a
periodically oscillating function of time. In the nonrigid pendulum analogy,
this corresponds that the motion of our system is turned from
\textquotedblleft running-phase modes\textquotedblright\ (Figure 1a) into
\textquotedblleft$\pi$-phase modes\textquotedblright\ (Figures 1b-1d) or from
a rotation into a vibration. Here the definitions of running-phase modes and
$\pi$-phase modes are the same as those for two weakly coupled BEC \cite{14}.

This critical behavior depends on $\Lambda_{c}=\Lambda_{c}\left[  \xi
_{2}\left(  0\right)  ,\xi_{3}\left(  0\right)  ,\varphi_{2}\left(  0\right)
=\frac{\pi}{2}\right]  $, as can be easily found from the energy conservation
constraint and the boundness of the tunneling energy in Eq. (\ref{12}). In
fact, the value $\left\langle \xi_{2}\left(  t\right)  \right\rangle <0$ is
inaccessible at any time\textbf{\ }if
\begin{align}
\Lambda &  >\Lambda_{c}=\frac{1}{3}-\xi_{2}\left(  0\right)  +\frac
{2\sqrt{1-9\xi_{3}^{2}\left(  0\right)  }}{9\xi_{2}\left(  0\right)
}\nonumber\\
&  -\frac{\left(  1+3\xi_{2}\left(  0\right)  \right)  \sqrt{\left(
2-3\xi_{2}\left(  0\right)  \right)  ^{2}-36\xi_{3}^{2}\left(  0\right)  }%
}{9\xi_{2}\left(  0\right)  }. \label{16}%
\end{align}
When $\varphi_{2}\left(  0\right)  =0$ the critical parameter is
\begin{align}
\Lambda_{c}  &  =\frac{1}{3}-\xi_{2}\left(  0\right)  -\frac{2\sqrt{1-9\xi
_{3}^{2}\left(  0\right)  }}{9\xi_{2}\left(  0\right)  }\nonumber\\
&  +\frac{\left(  1+3\xi_{2}\left(  0\right)  \right)  \sqrt{\left(
2-3\xi_{2}\left(  0\right)  \right)  ^{2}-36\xi_{3}^{2}\left(  0\right)  }%
}{9\xi_{2}\left(  0\right)  }, \label{17}%
\end{align}
and $\Lambda<\Lambda_{c}$ marks the regime of MQST. Specifically for $\xi
_{2}\left(  0\right)  =0.12,$ $\xi_{3}\left(  0\right)  =0.25,$ the critical
parameter is $\Lambda_{c}=-0.18.$ We take $\Lambda=-0.20$ and see MQST does
set in.

In Figure 2 we show solutions of Eqs. (\ref{11}) with initial conditions
$\xi_{2}\left(  0\right)  =0.12,$ $\xi_{3}\left(  0\right)  =0,$ $\varphi
_{2}\left(  0\right)  =0$ for parameters $\Lambda$ $=0.75,0.60,0.43$ and
$0.00,$ respectively. The left column shows again the time evolutions of the
population imbalance between $0$ and $\pm1$ components and indicates that
$\xi_{2}\left(  t\right)  $ is always a periodic function of time as $\Lambda$
decreases. For $\Lambda=\Lambda_{c}=0.43$ the population difference is self
locked to the initial value, which serves as another sign of the MQST
phenomenon. The right column shows the time evolution of the relative phase
between $0$ and $\pm1$ components and indicates that as $\Lambda$ decreases
$\varphi_{2}\left(  t\right)  $ is always a periodic function of time around
its mean value $\left\langle \varphi_{2}\left(  t\right)  \right\rangle =0$.
The dynamics corresponds to \textquotedblleft zero-phase
modes\textquotedblright. Moreover we observe that MQST occurs when
\begin{equation}
\Lambda<\Lambda_{c}=\frac{2}{3}-2\xi_{2}\left(  0\right)  , \label{18}%
\end{equation}
while\textbf{\ }for $\varphi_{2}\left(  0\right)  =\pi$ MQST never happens.

Therefore it is obvious that the spin-mixing dynamics depends on the relative
interaction parameter $\Lambda$ and is also sensitive to the initial
occupations and relative phases of the three components, which can be adjusted
by engineering Raman pulses \cite{20}. In practical experiments there are
usually two different ways to achieve MQST. In Figures 1 and 2, $\xi
_{2}\left(  0\right)  $ and $\varphi_{2}\left(  0\right)  $ are kept constants
while $\Lambda$ varies (by changing the geometry of condensates). On the other
hand, one can calibrate the initial values of the population imbalance
$\xi_{2}\left(  0\right)  $ with a fixed trap geometry (i.e., $\Lambda$
remains constant) and $\varphi_{2}\left(  0\right)  $ \cite{15}.

In order to further characterize the evolution of the system we
summarize the full dynamic behavior of Eq. (\ref{11}) in Figure 3
that shows the $\xi _{2}\left(  t\right)  $-$\varphi_{2}\left(
t\right)  $ phase portrait with constant energy contour. The
distinction between the two dynamic regimes -- nonlinear Josephson
tunneling and MQST -- becomes more apparent in Figure 3. In the
regime of Josephson oscillation the dynamic variables follow a
closed phase space trajectory, while in the self-trapping regime
they follow an open trajectory with an unbounded phase.

We are inspired by the great expectation that tuning the contact interaction
between cold atoms close to zero by Feshbach resonance \cite{21} will make the
MDDI more prominent or even the dominant interaction \cite{10}. Based on the
experimental values of the $s$-wave scattering lengths for $^{23}$Na,
$a_{0}=\left(  50.0\pm1.6\right)  a_{B}$ and $a_{2}=\left(  55.0\pm1.7\right)
a_{B}$ with $a_{B}$\ being the Bohr radius \cite{23}, the ratio of
coefficients $c_{d}$ and $\left\vert c_{2}\right\vert $ can be shown to be
$0.007$, while for $^{87}$Rb($a_{0}=101.8a_{B}$ and $a_{2}=100.4a_{B}$) it is
$0.1$ \cite{6}. Evaluation with a simple variational wave function \cite{24}
gives $U_{2}N\simeq6nK$ for a sodium condensate up to $N=5\times10^{6}$ atoms
confined in an optical dipole trap with a very small trapping volume
($10^{-8}$ $cm^{-3}$) \cite{1,22}. On the other hand, as argued in Refs.
\cite{4,6,10,17,18}, both the sign and the magnitude of the MDDI $U_{d}$ can
be greatly tuned with trapping geometry. As has been shown \cite{6},
$U_{d}/U_{2}$ depends on a monotonically increasing function of the condensate
aspect ratio $\kappa$, bounded between $-1$ and $2$. This clearly shows the
possibility of adjusting the MDDI very close to spin exchange interaction,
i.e. $U_{d}/U_{2}\simeq1$. At this point our parameter $\Lambda$ may take
values in a large scale and the manifestation of MQST is within reach with
current technologies. We point out although for all of the alkali-metal atomic
condensations the MDDI is rather weak compared to the contact potential, the
experimental achievement of BEC with transition-metal chromium $^{52}Cr$
provides us the hope because the MDDI here is 36 times stronger than that of
alkali atoms. For these reasons, a degenerate quantum gas with adjustable
long- and short-range interactions can be experimentally realized in near
future. By altering the strengths of two kinds of interactions and the initial
conditions, the nonlinear tunneling dynamics of dipolar spinor BEC
consequently sustains a self-maintained population imbalance: a novel MQST
effect. Such a scheme would avoid the difficulty\ of realizing experimentally
the double-well magnetic trap.

\section{Summary}

In summary we have described a semiclassical treatment of the spin-1 dipolar
spinor condensates.\ As a result of the conservation of atom numbers and total
hyperfine spin of the condensate, the classical equations of motion are
derived and discussed in a similar way as in double-well BJJ. It is
demonstrated that spontaneous magnetization and spin-mixing dynamics depend
on\textbf{\ }both\textbf{\ }the spin-exchange interaction $U_{2}$ and the MDDI
$U_{d}$ through the ratio $\frac{3U_{d}}{2\left\vert U_{2}-U_{d}\right\vert
N}.$ The initial population imbalance, the relative phase among the three
components of the condensate as well as the total hyperfine spin of the system
all play important roles in the semiclassical dynamics. Finally we have
indicated the possibility of using dipolar spinor condensate as a platform for
practical manipulation of MQST and Josephson oscillation.

\section{Acknowledgment}

This work was supported by National Natural Science Foundation of China under
Grant Nos.10475053, 10175039 and 90203007.

* corresponding author: ybzhang@sxu.edu.cn

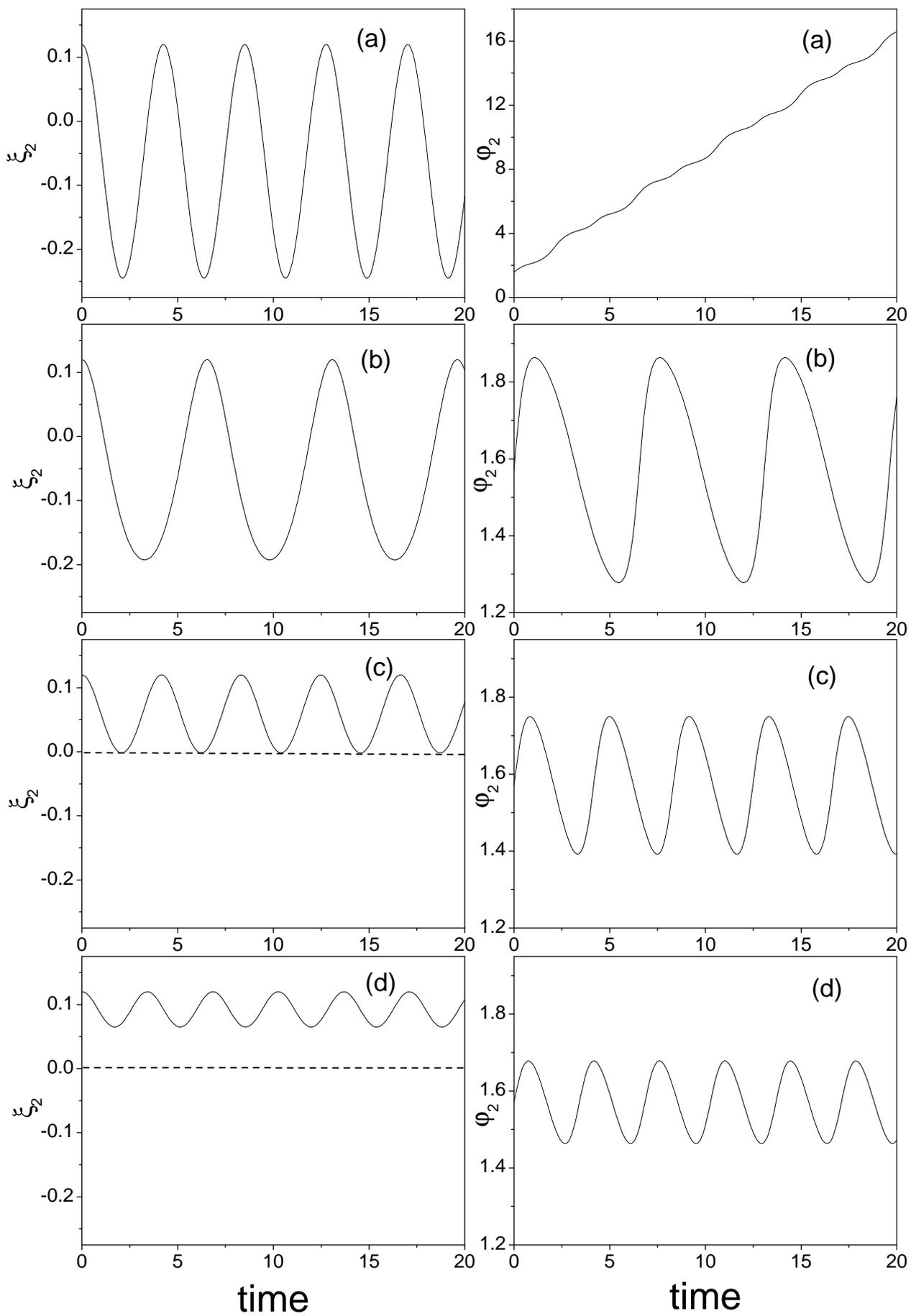
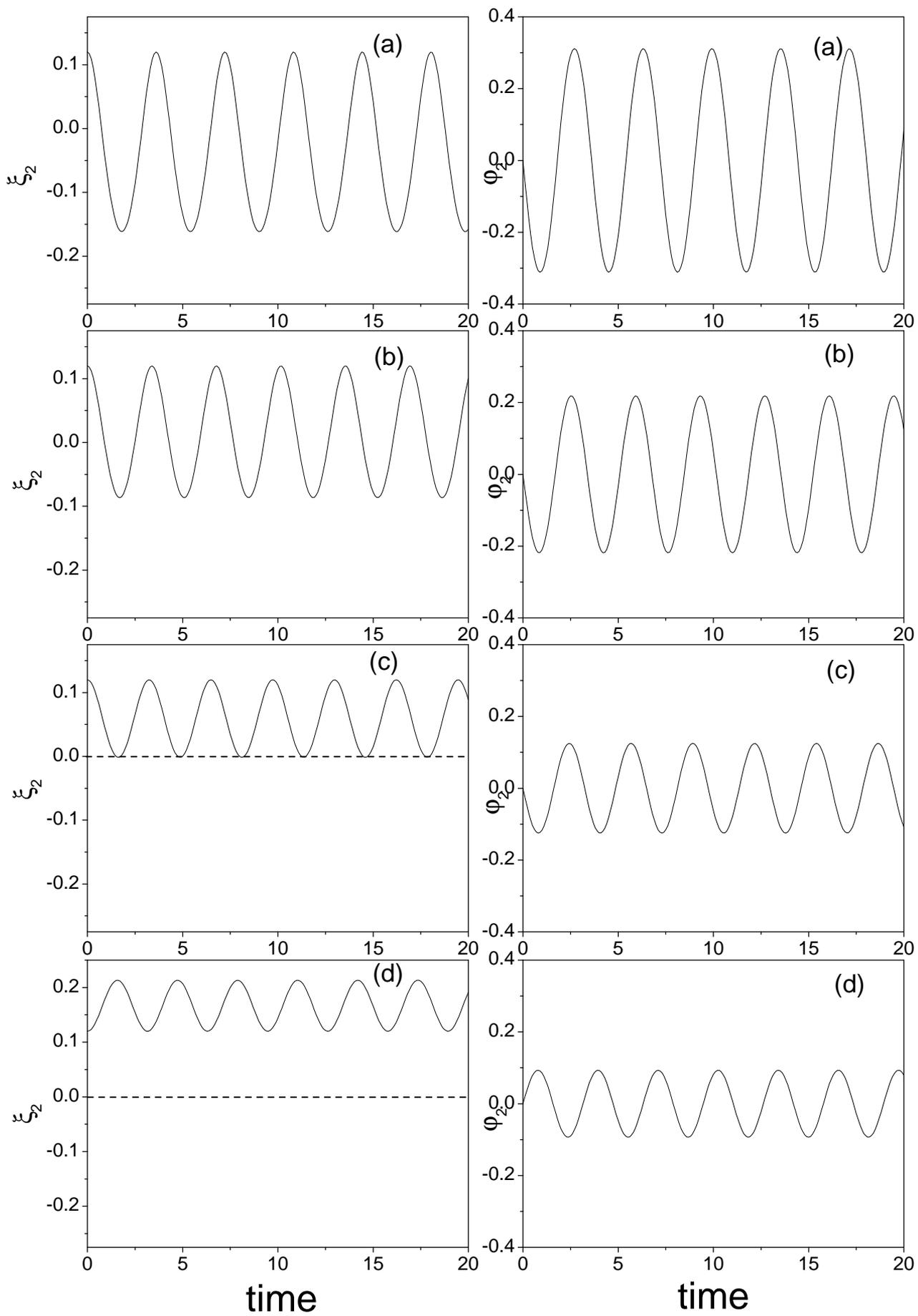
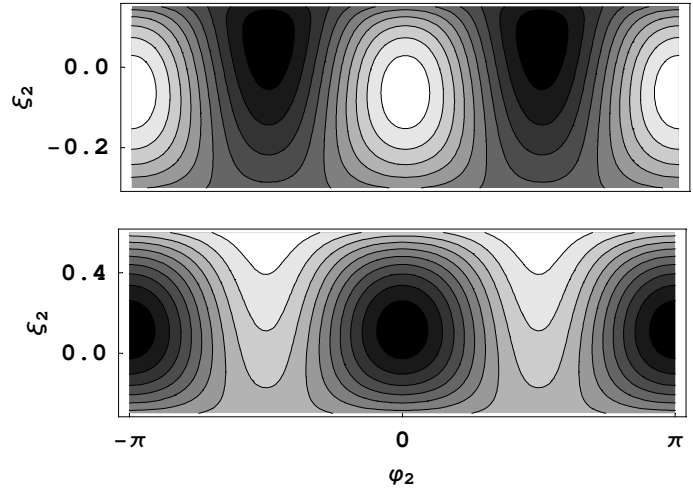
\end{document}